\documentclass[letterpaper,10 pt, conference]{ieeeconf}  

\IEEEoverridecommandlockouts                              

\usepackage{xcolor}

\usepackage[fleqn]{amsmath}
\usepackage{amssymb,amsfonts,amsthm} 
\usepackage{tabularx}
\usepackage{mathabx}
\usepackage{graphicx,url} 
\usepackage{bm}
\usepackage{float}
\usepackage{makecell}
\let\labelindent\relax
\usepackage{enumitem}
\usepackage{subcaption}
\usepackage{wrapfig}
\usepackage{algorithm,algorithmicx} 
\usepackage{algpseudocode} 
\usepackage{multirow}
\usepackage{diagbox}

\usepackage{cite}
\usepackage{soul}  
\usepackage[bookmarks=true]{hyperref}
\hypersetup{
    colorlinks=true,
    pdfborderstyle={/S/U/W 1},
    linkcolor=blue,
    filecolor=magenta,      
    urlcolor= blue,
    linkbordercolor=red,
}
\newcommand\rurl[1]{%
  \href{http://#1}{\nolinkurl{#1}}%
}

\usepackage{comment}
\usepackage{accents}

\makeatletter
\def\cl@part {\@elt {chapter}}
\makeatother

\usepackage{cleveref}

\crefname{equation}{}{} 
\crefname{lemma}{Lemma}{Lemmas}
\crefname{theorem}{Theorem}{Theorems}
\crefname{table}{Table}{Tables}
\crefname{figure}{Fig.}{Figs.}
\crefname{remark}{Remark}{Remarks}
\crefname{assumption}{Assumption}{Assumptions}
\crefname{section}{Section}{Sections}
\crefname{definition}{Definition}{Definitions}
\crefname{algorithm}{Algorithm}{Algorithms}
\crefname{proposition}{Proposition}{Propositions}
\crefname{appendix}{Appendix}{Appendices}

\usepackage{accents}

\theoremstyle{definition}  
\theoremstyle{definition} 

\theoremstyle{remark}  
\newtheorem{remark}{Remark}

\makeatletter
\renewcommand*\env@matrix[1][\arraystretch]{%
  \edef\arraystretch{#1}%
  \hskip -\arraycolsep
  \let\@ifnextchar\new@ifnextchar
  \array{*\c@MaxMatrixCols c}}
\makeatother

\usepackage{commath}

\def \onetoN#1 {1,\dots,#1}

\usepackage{textcomp}
\makeatletter
\newcommand\fs@spaceruled{\def\@fs@cfont{\bfseries}\let\@fs@capt\floatc@ruled
  \def\@fs@pre{\vspace{3mm}\hrule height.8pt depth0pt \kern2pt}%
  \def\@fs@post{\kern2pt\hrule\relax\vspace{-4mm}}%
  \def\@fs@mid{\kern2pt\hrule\kern2pt}%
  \let\@fs@iftopcapt\iftrue}
\makeatother

\renewcommand{\arraystretch}{1.5}

\author{Adam Hallmark and Pan Zhao,~\IEEEmembership{Member,~IEEE} 
\thanks{A. Hallmark and P. Zhao are with the Department of 
Aerospace Engineering and Mechanics, University of Alabama, Tuscaloosa, AL 35487, USA. Email: {\tt\small alhallmark@crimson.ua.edu}, {\tt\small pan.zhao@ua.edu}. }
}

\begin{document}

\title{\LARGE \bf Improving INDI for Input Nonaffine Systems via Learning-Based \\ Nonlinear Control Allocation\\

\thanks{This research was financially supported in part by the Alabama Space Grant Consortium, NASA Training Grant NNH24ZHA003C}
}

\maketitle
\thispagestyle{empty}
\pagestyle{empty}

\begin{abstract}
This paper first demonstrates that applying standard incremental nonlinear dynamic inversion (INDI) with incremental control allocation (ICA) to input nonaffine systems relies on an untenable linear approximation of the actuator model. It then shows that avoiding this issue,  while retaining the static control allocation paradigm, generally requires solving a nonlinear programming (NLP) problem.
To address the associated online computational challenges, the paper subsequently presents a supervised learning–based approach. Numerical experiments on an example problem validate the identified limitations of standard INDI + ICA for input nonaffine systems, while also demonstrating that the proposed learning-based method provides an effective and computationally tractable alternative.

\end{abstract}



\section{Introduction}\label{sec:introduction}

Nonlinear dynamic inversion (NDI), also known as feedback linearization, is a control design technique  for a class of nonlinear systems that uses feedback to linearize the system response with respect to a virtual input \cite{khalil_nonlinear_2002, Slotine_Li_1991, enns1994_dynamic_inversion_evolving_method_fc_design}. This enables the application of linear control design techniques to shape the response of controlled variables for nonlinear systems. To achieve exact feedback linearization, the system model must be known exactly. For systems with complex dynamics or subject to disturbances, accurate dynamic models may be difficult to obtain. This practical difficulty has motivated the development of incremental NDI (INDI), which is a modification of standard NDI that incorporates sensor measurements to reduce model dependence. INDI has become a popular method for the control of aerospace systems \cite{smeur2018cascaded_INDI_for_MAV_dist, sieberling2010robust_FC_INDI, lu2016aircraft_ftc_using_INDI, di2016modeling_INDI_tiltrotor}. See the survey \cite{steinert2025_INDI_survey1_CJA} for references to additional applications.

Overactuated systems are generally defined as systems that have more inputs than controlled variables \cite{CA_for_overactuated_sys}. Control allocation (CA) is an approach used to manage the distribution of actuator effort for such systems \cite{johansen2013control_allocation_a_survey}. When INDI is applied to overactuated input nonaffine systems, the typical derivation leads naturally to a linear CA problem. The resulting formulation is often called the incremental control allocation (ICA) problem, and it has been applied to aerospace systems \cite{pfeifle2023minimum, matamoros2018_INCA_for_ICE}. This approach involves local linearization of the actuator model so that the allocation problem can be cast as a linear one. In prior literature, it has been claimed that the use of this local linear approximation is actually an attractive feature of INDI over standard NDI that enables application to nonaffine systems. For example, one of the earliest works in this area states that ``The problem of applying DI to a system with a nonaffine control mapping has also been eliminated" \cite{bacon2000_early_INDI}. While the survey \cite{steinert2025_INDI_survey1_CJA} does mention that the local linearization is only an approximation, it also states that one of the key advantages of INDI over NDI is that it ``is suitable for input non-affine systems".

While local linearization of the actuator model does enable the use of efficient linear CA methods for computing the control increment, the error involved in the approximation causes a direct degradation of the feedback linearization performance. Further, the assumption that would be necessary to ensure small degradation--a small control increment--conflicts with one of the assumptions commonly used to derive the INDI control law. To avoid the errors associated with local linearization of the actuator model while retaining the static control allocation approach, a nonlinear control allocation (NCA) problem must be solved.

In this paper, we explicitly and precisely identify the problem with applying the standard INDI formulation with ICA to nonaffine systems, and we argue that the problem is fundamental. We then derive a control law and allocation scheme for nonaffine systems that is analogous to INDI but avoids the standard local linearization. Our formulation requires the solution of an NCA problem to compute the control input, and we propose a supervised learning-based approach to solve the resulting NCA problem. Simulation results  validate the identified issues with 
the standard INDI approach for input nonaffine systems, and also demonstrate the potential of the proposed learning-based approach as an effective
and computationally-tractable alternative for solving NCA problems.

\section{The Problem with Applying Standard INDI + LCA to Input Nonaffine
Systems}\label{sec:standard_indi_problems}

We begin by following a standard INDI control law derivation, similar to that of \cite{bacon2000_early_INDI, steinert2025_INDI_survey1_CJA}. Consider the following nonaffine system:
\begin{equation}
    \dot{x} = F(x,u) = f(x) + g(x,u),
    \label{eq:nonaffine_dyn}
\end{equation}
where $x \in \mathbb{R}^n$ and $u \in \mathbb{R}^m$ with $m>n$. We assume that the system \cref{eq:nonaffine_dyn} is fully input–state feedback linearizable with a well-defined vector relative degree, such that each state component can be independently linearized and assigned its own virtual control input. This simplifies our analysis and does not affect the main results. We pursue feedback linearization via INDI in such a case. The Taylor series expansion of \cref{eq:nonaffine_dyn} around a particular state $x_0$ and input $u_0$ is given by
\begin{equation}
\begin{split}
    \dot{x} = F(x_0, u_0) + \frac{\partial F}{\partial x}(x_0, u_0)\Delta x + \frac{\partial F}{\partial u}(x_0, u_0) \Delta u \\
    + \mathcal{O}(\| \Delta x \|^2) + \mathcal{O}(\| \Delta u \|^2) + \mathcal{O}(\| \Delta x \| \| \Delta u \|),
\end{split}
\label{eq:taylor_exp_nonaffine}
\end{equation}
where $\Delta x \coloneq x - x_0$ and $\Delta u \coloneq u - u_0$. In the typical INDI derivation, only the first-order terms in \cref{eq:taylor_exp_nonaffine} are included and the heuristic ``time scale separation principle" \cite{matamoros2018_INCA_for_ICE} is invoked to argue that the approximation $ \Delta x \approx 0$ is valid since the control inputs change faster than the states \cite{sieberling2010robust_FC_INDI}. In fact, the continuity of $x(t)$ is sufficient to establish that $\Delta x \approx 0$ is a valid approximation when $x_0$ and $x$ are the states at some time $t_0$ and at some later time $t = t_0 + \Delta t$, respectively, with $\Delta t > 0$ being small. 
Thus, the second, fourth, and sixth terms on the right hand side of \cref{eq:taylor_exp_nonaffine} can be reasonably neglected for sufficiently small $\Delta t$. In our developments, $\Delta t$ corresponds to the sample time of the realized control system.

\subsection{Review of INDI for input affine systems}
In the control-affine case, we have in \cref{eq:nonaffine_dyn} that
$$
    g(x,u) = B(x)u,
$$
which implies that
$$
    \frac{\partial^i F}{\partial u^i}(x_0, u_0) = 0, \quad \forall i > 1.
$$
This means that the fifth term on the right-hand side of \cref{eq:taylor_exp_nonaffine} is precisely zero in the control-affine case. Let us proceed using the identity $\mathcal{O}(\| \Delta u \|^2) = 0$ to derive the control law. We will reintroduce this collection of terms subsequently in discussing the case of input nonaffine systems. For ease of notation, let us define
$$
    B_0 \coloneq \frac{\partial F}{\partial u}(x_0, u_0) = \frac{\partial g}{\partial u}(x_0, u_0),
$$
where the equality on the right follows from \cref{eq:nonaffine_dyn}.
We can now write the approximate relation
\begin{equation}
    \dot{x} \approx F(x_0, u_0) + B_0 \Delta u.
    \label{eq:approx_affine}
\end{equation}
In typical INDI derivations/applications, the term $F(x_0, u_0)$ is replaced by $\dot{x}_0$ which is directly measured or estimated using on-board sensors, meaning that we do not need to know the drift term $f(x)$ in order to compute the control input. This control design choice makes the standard INDI naturally robust in the sense that the controller is less model-dependent and any disturbances injected into the system will be captured in $\dot{x}_0$. The use of estimated or measured $\dot{x}_0$ is also why INDI is often referred to as a sensor-based control method \cite{wang2019stability}.

Let us introduce the so-called virtual control input $\nu$. If we could design a control law for $\Delta u$ that achieves $\dot{x} \approx \nu$, then we would be able to arbitrarily specify the approximate linear closed-loop dynamics for $x$ with suitable choice of feedback law for $\nu (x)$, e.g., $\nu(x) = -K(x-x_c)$ for tracking control, where $x_c$ is the commanded state. Replacing $F(x_0, u_0)$ with the sensor estimate $\dot{x}_0$ and equating the right hand side of \cref{eq:approx_affine} with $\nu(x)$, which yields $\dot{x} \approx \nu(x)$, we have
\begin{equation}
    \nu(x) = \dot{x}_0 + B_0 \Delta u.    \label{eq:implicit_relation_standard_INDI}
\end{equation}
Assume that we do indeed select $\nu(x) = -K(x-x_c)$ as the virtual control law. Then, to compute the control increment $\Delta u$, we need to solve at each time step
\begin{equation}
    -K(x-x_c) - \dot{x}_0 = B_0 \Delta u.    \label{eq:linear_incremental_CA}
\end{equation}
Any constrained or unconstrained linear control allocation method \cite{johansen2013control_allocation_a_survey} can be used to determine a $\Delta u$ satisfying \eqref{eq:linear_incremental_CA}, e.g., via solving a quadratic programming (QP) problem:
\begin{equation}
\label{eq:QP_alloc_general_form}
\begin{aligned}
    \min_{\Delta u}& \quad \Delta u^T Q \Delta u\\
    \textrm{s.t.}& \quad \text{condition \cref{eq:linear_incremental_CA}},
\end{aligned}
\end{equation}
where $Q \geq 0$. Note that the constrained approach of QP-based linear CA allows for the inclusion of inequality constraints on $\Delta u$ (actuator limits) in \cref{eq:QP_alloc_general_form}. The actual control input supplied to the system is then determined by $u = u_0 + \Delta u$, where $u_0$ is the input from the previous sample instant.

\subsection{Issues with standard INDI for input nonaffine systems} \label{sec:indi-issue-nonaffine}
In order to derive the implicit incremental control law in \cref{eq:linear_incremental_CA} which achieves the approximate feedback linearization $\dot{x} \approx \nu(x)$, we first assumed that $\mathcal{O}(\| \Delta u \|^2)$ from \cref{eq:taylor_exp_nonaffine} was equal to zero. Recall that this equality is actually guaranteed to be true in the input affine case. In the case of input nonaffine systems, let us assume that $\Delta u$ is selected according to \cref{eq:linear_incremental_CA} and then investigate whether or not the approximation $\mathcal{O}(\| \Delta u \|^2) \approx 0$ is reasonable (which, in turn, tells us whether or not we will actually achieve $\dot{x} \approx \nu(x)$ if we use \cref{eq:linear_incremental_CA}). Note that {\it the approximation of $\mathcal{O}(\| \Delta u \|^2) \approx 0$ is justified only when $\| \Delta u \| \ll 1$}. Immediately, we can see from \cref{eq:linear_incremental_CA} that a large step change in $x_c$ from one sample to the next would generally imply a large $\| \Delta u \|$, invalidating the approximation which was used to arrive at \cref{eq:linear_incremental_CA}. Even in the case where the virtual control is designed for regulation of $x$ to a fixed desired steady-state value, i.e., $\nu(x) = -K(x-x_{ss})$, we can see from \cref{eq:linear_incremental_CA} that a large change in the disturbance (observed through measurement of $\dot{x}_0$) from one sample to the next would also generally imply a large $\| \Delta u \|$. Thus, {\it abrupt changes in either the virtual control or the potential disturbance are generally unacceptable for achieving accurate feedback linearization when applying the standard INDI to input nonaffine control systems}. Note that this is also true when $\text{dim}(x) = \text{dim}(u)$, i.e., in the non-overactuated case. In summary, there is no intrinsic mechanism in \cref{eq:linear_incremental_CA} to ensure sufficiently small control increments. 
This may substantially limit practical applicability, as feedback linearization is often used to design tracking controllers for rapidly-varying reference commands, while sensor-based approaches are employed to enable the rejection of (large) disturbances.

Even if we were to filter the reference command, i.e., enforce continuity of $x_c(t)$, and establish bounds on the time derivatives of $x_c(t)$ and of any possible disturbance $d(t)$, we would still need to incorporate these bounds in selecting the appropriate sample time for controller implementation to ensure that $\| \Delta u \| \ll 1$ always holds. This requirement may lead to a control rate that is infeasible in practice, depending on the magnitude of the established bounds. Moreover, the objective of keeping $\| \Delta u \|$ small is in direct conflict with the heuristic “time-scale separation principle” commonly invoked in deriving the INDI control law. In the case of input-output feedback linearization, similar conclusions apply.

In the next section, we show how to derive the control law and allocation scheme for input nonaffine systems, which addresses the aforementioned problems.

\section{Modified INDI + Allocation Scheme for Input Nonaffine
Systems}\label{sec:indi_inspired_nonaffine}
To address the issues of standard INDI for input nonaffine systems identified in \cref{sec:indi-issue-nonaffine}, we will develop a control law and allocation scheme for input nonaffine systems that is similar in spirit to the standard INDI formulation, while completely avoiding the assumption $\mathcal{O}(\| \Delta u \|^2) \approx 0$. The control law for the actual input will be expressed in implicit form, similar to \cref{eq:linear_incremental_CA}, but will require the solution of a nonlinear control allocation (NCA) problem to compute the control input rather than a linear CA problem.

Consider again the system \cref{eq:nonaffine_dyn}. Instead of writing the full Taylor expansion in both arguments $x$ and $u$, consider the partial Taylor expansion of the dynamics with respect to only $x$ around a particular state $x_0$ while treating $u$ as a parameter:\begin{equation}
    \dot{x} = F(x_0, u) + \frac{\partial F}{\partial x}(x_0, u) \Delta x + \mathcal{O}(\| \Delta x \|^2).
    \label{eq:partial_taylor_nonaffine}
\end{equation}
Following the same arguments from \cref{sec:standard_indi_problems} related to the continuity of $x(t)$, we can reasonably make the assumption that $\Delta x \approx 0$ for sufficiently small sample time $\Delta t$ of the eventual control implementation. Thus, we can reasonably neglect the second and third terms on the right hand side of \cref{eq:partial_taylor_nonaffine}, leading to
\begin{equation}
    \dot{x} \approx F(x_0, u) = f(x_0) + g(x_0, u).
    \label{eq:approx_xdot_partial_taylor}
\end{equation}
We would like to select a control input $u$ to achieve the approximate feedback linearization with respect to the virtual input: $\dot{x} \approx \nu$. To this end, equate the right hand side of \cref{eq:approx_xdot_partial_taylor} with the desired dynamics $\nu(x)$, which yields $\dot{x} \approx \nu(x)$, and we have
\begin{equation}
    \nu(x) = f(x_0) + g(x_0, u).
    \label{eq:imp_rel_no_meas_nonaffine}
\end{equation}
Observe that, unlike \cref{eq:implicit_relation_standard_INDI}, equation \cref{eq:imp_rel_no_meas_nonaffine} does not involve the sensor estimate $\dot{x}_0$ and also depends on knowledge of the drift term $f(x)$. To include sensor feedback of the state derivative and retain the spirit of standard INDI control, we can replace $f(x_0)$ by $\dot{x}_0 - g(x_0, u_0)$\footnote{Notice that $x_0$ and $u_0$ satisfy the dynamics \cref{eq:nonaffine_dyn}, i.e., $\dot x_0 = f(x_0)+g(x_0,u_0)$.} and obtain the control law for $u$ in implicit form:
\begin{equation}
    \nu(x) = \dot{x}_0 - g(x_0, u_0) + g(x_0, u).
    \label{eq:imp_rel_meas_nonaffine}
\end{equation}
If we rearrange \cref{eq:imp_rel_meas_nonaffine} and recall that $u$ is of a higher dimension than the state $x$, we obtain the NCA problem in a typical form \cite{johansen2013control_allocation_a_survey}:
\begin{equation}
    \underbrace{\nu(x) - \dot{x}_0 + g(x_0, u_0)}_{\coloneq \mu_{des}} = g_u(u) \coloneq g(x_0, u),
    \label{eq:nca_nonaffine}
\end{equation}
where $\mu_{des}$ is the desired \textit{generalized input} and can be thought of as the desired acceleration due to input. In general, we can attempt to solve the NCA problem during each sample period using the nonlinear programming (NLP) formulation as follows:
\begin{equation}
\label{eq:NLP_alloc_general_form}
\begin{aligned}
    \min_{u}& \quad J(u) \\
    \textrm{s.t.}& \quad \mu_{des} = g_u(u),
\end{aligned}
\end{equation}
where $J(u)$ represents the static secondary control objective such as a power usage or ``effort" proxy, e.g., $J(u) = u^T Q u$. Inequality constraints on the input $u$ (actuator limits) can also be added to \cref{eq:NLP_alloc_general_form}, similar to how the constrained linear CA approach may be used to solve \cref{eq:linear_incremental_CA} in the presence of actuator limits. Since \cref{eq:linear_incremental_CA} or \cref{eq:NLP_alloc_general_form} would be solved in a sampled fashion, one can also incorporate actuator rate limits by limiting the change in control $\Delta u$ from the control $u_0$ at the previous sample instant using $\Delta u_{max} = \dot{u}_{max} \Delta t$ in combination with the absolute limits on the control inputs. 

\begin{remark}
\label{remark:ca_relaxation}
When rate or absolute limits on actuators are included, one must consider that \cref{eq:linear_incremental_CA} or \cref{eq:NLP_alloc_general_form} may sometimes be infeasible. In such cases, the equality constraint is typically relaxed using a penalized slack variable. Though this approach generally makes the computational problem feasible, a nonzero slack at any iteration of the control loop will introduce some additional pointwise error in the approximate relation $\dot{x} \approx \nu(x)$. We can attempt to address this potential issue by limiting/filtering the signal $\nu(x)$, or, in the case of multiple controlled variables, by prioritizing the minimization of equality constraint errors for \textit{certain channels} of the state which are most important for safety, for example \cite{johansen2013control_allocation_a_survey}.
\end{remark}

\begin{remark}
    \label{remark:actuator_model_dependence}
    In some applications, the implicit control law \cref{eq:nca_nonaffine} may be used in an inner loop while other control design methods are utilized for the higher-level controllers, so that the function $g(x,u)$ may depend additionally on variables representing integrals of the states in \cref{eq:nonaffine_dyn}, or on other time-varying parameters. Provided that these variables/parameters have continuous trajectories and can be measured or estimated, equation \cref{eq:nca_nonaffine} can be straightforwardly modified to include dependence of $g$ on the additional variables/parameters at the previous sample instant.
\end{remark}

The NLP \cref{eq:NLP_alloc_general_form} is generally nonconvex due to the presence of a nonaffine equality constraint. Accordingly, direct NLP formulations of the NCA problem are often avoided in real-time control applications, as solvers for nonconvex problems typically lack the computational reliability and efficiency of convex optimization methods. In the next section, we propose an approach to address this issue.

\section{Learning-Based Nonlinear Control Allocation}\label{sec:proposed_learning_approach}

In this section, we describe a learning-based approach to provide approximate online solutions to the NCA problem in \cref{eq:nca_nonaffine} with fast evaluation. We explain how our proposed learning-based approach is distinct from prior work, e.g., \cite{khan2024_nonlin_CA_learning_based_approach, huan2018_ECC_NCA_using_deep_AE_NN}, and argue that our proposed approach offers certain benefits over existing learning-based methods for general NCA. 

Consider \cref{eq:nonaffine_dyn} and \cref{eq:nca_nonaffine}, and let us reintroduce the so-called generalized input $\mu$:
\begin{equation}
    \mu \coloneq g(x,u).
    \label{eq:mu_definition}
\end{equation}
We can interpret $\mu$ as an intermediate variable that directly represents the instantaneous affine effect of $u$ (with unity gain) on the state derivative $\dot{x}$ for a particular instantaneous value of $x$. In the case of motion control of aerospace vehicles, $\mu$ often has the physical meaning of the control-produced vehicle accelerations. In view of \cref{eq:nca_nonaffine} and \cref{eq:mu_definition}, the NCA problem can be understood as: given a desired $\mu$ and previous state $x_0$, determine an optimal $u^*$ such that \cref{eq:mu_definition} holds with $\mu = \mu_{des}$ and $x = x_0$. The optimality here is with respect to $J(u)$ in \cref{eq:NLP_alloc_general_form}. Let $x \in \mathcal{X}$, $u \in \mathcal{U}$, and $\mu \in \mathcal{M}$ where $\mathcal{X} \subseteq{\mathbb{R}^n}$ is some domain of interest containing the origin, $\mathcal{U} \subseteq \mathbb{R}^m$ is the admissible set of inputs, and $\mathcal{M} \subseteq \mathbb{R}^n$ is the image of $\mathcal{X} \times \mathcal{U}$ under the mapping $g(x,u)$ in \cref{eq:mu_definition}. Let $\mathcal{M}_0$ denote the image of $\mathcal{U}$ under the mapping $g(x_0, u)$ for a particular state $x_0$. The set $\mathcal{M}_0$ is often referred to as the {\it attainable moment set} in the aerospace control literature, with the name coming from consideration of the aircraft attitude control problem\cite{Durham1993_constrained_control_allocation, Durham1993_CCA_3M_problem}. The set $\mathcal{M}$ is thus the union of pointwise attainable moment sets across all $x_0 \in \mathcal{X}$.

The function $g(x,u)$ may not depend explicitly on all $n$ components of the state $x$. As discussed in \cref{remark:actuator_model_dependence}, $g(x,u)$ may also depend on additional variables not included in $x$. Let $x_g$ denote the $n_g$-dimensional vector containing all variables upon which $g(x,u)$ depends explicitly, aside from $u$. Further, with a slight abuse of notation, let us from now on refer to $g$ as depending explicitly on the arguments $(x_g,u)$, rather than $(x,u)$. With the subscript $L$ referring to ``Learning", define $x_L \coloneq [x_{g,0}^T, \mu_{des}^T]^T$ and $y_L \coloneq u^*$, where $u^*$ is the corresponding optimal input vector with respect to $J(u)$ in \cref{eq:NLP_alloc_general_form}. Let the dimension of $x_g$ be $n_g$ and let the set $\mathcal{X}_g \subseteq \mathbb{R}^{n_{g}}$ denote the region of interest for $x_g$. Let the dimension of $x_L$ be  $n_{x_L}$, where $n_{x_L} = n_g + n$. Define the set $\mathcal{X}_A \subseteq \mathbb{R}^{n_{x_L}}$ via the property:
 $$
 [x_{g,0}^T, \mu_{des}^T]^T \in \mathcal{X}_A \implies \exists u \in \mathcal{U}: g(x_{g,0}, u) = \mu_{des},
 $$
and note that $\mathcal{X}_A \neq \mathcal{X}_g \times \mathcal{M}$. To see this, observe that for a particular $x_{g,0} \in \mathcal{X}_g$, the set of achievable $\mu$ is the pointwise attainable moment set $\mathcal{M}_0$ corresponding to $x_{g,0}$, not the entire union $\mathcal{M}$. The set $\mathcal{X}_A$ is simply the set of all pairs $(x_{g,0},\mu_{des})$ such that $\mu_{des}$ can be achieved under the input constraints when $x_g = x_{g,0}.$

\subsection{Learning problem setup and training data generation}\label{sec:data-generation-learning}
The key idea of the learning-based approach is to learn the inverse map $\mathcal{X}_A \to \mathcal{U}$. We propose to learn this mapping using neural networks (NNs) trained via a supervised learning approach where the labels are generated through offline solution of many NLP problems.

We require $N$ samples from the set $\mathcal{X}_A$ along with the $N$ corresponding optimal input vectors $y_L$, obtained by numerically solving the NLP problem \cref{eq:NLP_alloc_general_form}. Denote the $N \times n_{x_L}$ array of input data points by $X$, and the $N \times m$ array of labels, or output data, by $Y$. The learning problem is formulated as a regression task using the labeled data $(X, Y)$ and a chosen NN architecture. The resulting trained model takes the form:
\begin{equation}
    \hat{u}^* = \phi_{NN}([x_{g,0}^T, \mu_{des}^T]^T; \theta_{NN}),
    \label{eq:regression_model}
\end{equation}
where $\hat{u}^*$ denotes the predicted optimal control input and $\theta_{NN}$ represents the network parameters.

To directly sample the points $x_L^i, i = 1, \dots, N$ from the set $\mathcal{X}_A$, an explicit description of $\mathcal{X}_A$ is required. In general, such a description cannot be easily obtained since $g(x_g,u)$ is nonlinear. We propose the following heuristic method under the assumption that $\mathcal{U}$ is a hyperrectangle:
\begin{enumerate}
    \item Define the region of interest $\mathcal{X}_g$ and obtain a coarse grid of $N_x$ points over $\mathcal{X}_g$. Next, obtain the vertices of $\mathcal{U}$ using the Cartesian product of the component-wise lower and upper bounds, which yields $N_V = 2^m$ vertices;
    \item Obtain the Cartesian product of the $N_x$ sampled points in $\mathcal{X}_g$ and the $N_V$ vertices of $\mathcal{U}$, which yields $N_p = N_x \times N_V$ points in $\mathcal{X}_g \times \mathcal{U}$;
    \item Evaluate $g(x_g^i,u^i)$ for $i = 1, \dots, N_p$ which yields the points $\mu^i, i = 1, \dots, N_p$ in the set $\mathcal{M}$. Form the corresponding $N_p$ points in $\mathcal{X}_A$ as $x_A^i = [(x_g^i)^T, (\mu^i)^T]^T, i = 1, \dots, N_p$;
    \item Compute the convex hull of these $N_p$ points in $\mathcal{X}_A$. Denote the hull by $P_{A}$, and denote the number of vertices of $P_A$ by $N_{V,P}$;
    \item Sample $N_{s}$ points from within $P_{A}$ and add these to the vertices of $P_A$ to obtain $N_{init} = N_s + N_{V,P}$ sampled points;
    \item For each of the $N_{init}$ sampled points, attempt to solve the NLP problem \cref{eq:NLP_alloc_general_form} numerically. Reject the samples for which a solution to the corresponding NLP problem cannot be obtained, which will result in $N \leq N_{init}$ pairs of labeled data $(X,Y)$ for training.
\end{enumerate}

In the case that $N$ is smaller than desired after this procedure, one can simply increase $N_s$ until $N$ reaches some desired threshold. Note that the reason that the NLP problems may sometimes be \textit{infeasible} is that $[x_{g,0}^T, \mu_{des}^T]^T \in P_A$ does not imply that $[x_{g,0}^T, \mu_{des}^T]^T \in \mathcal{X}_A$. In addition,  there may be some points in $\mathcal{X}_A$ that are not in $P_A$. The former problem only causes wasted computational effort, whereas the latter leads to the possibility that certain regions of $\mathcal{X}_A$ will not be covered by the training data. We do not address the coverage issue in this paper, but we note that one could use an exploratory sampling+rejection scheme to try and extend the covered region of $\mathcal{X}_A$ outside of $P_A$.

\begin{remark}
    \label{remark:nonsmooth_map}
     One potential concern in the learning problem formulation is that the mapping $\mathcal{X}_A \to \mathcal{U}$ may be non-smooth or even discontinuous. If the training data exhibit sharp local jumps, for example as indicated by large output variations for nearby samples in $X$, then a single global regressor may be inadequate. In such cases, specialized network architectures, such as piecewise or gated neural network architectures, may be more appropriate.
\end{remark}

\begin{remark}
    \label{remark:global_min_issue}
    We assume that the feasible NLP problems required for training-data generation can be solved reliably in an offline setting. The equality constraint in \cref{eq:nca_nonaffine} is enforced up to a prescribed numerical tolerance, and sampled points for which the NLP solver fails to converge are rejected. Such rejection may arise either from true infeasibility or from numerical issues associated with the nonconvex optimization problem, including convergence to poor local minima. Because these NLPs are solved offline, more computationally intensive optimization strategies may be employed, such as multi-start or basin-hopping methods. In the case of a nonsmooth mapping $g(x_g,u)$--such as when when the actuator model is derived from interpolated CFD or flight-test data--gradient-free optimization methods such as particle swarm optimization (PSO) or genetic algorithms (GA) may also be considered.
\end{remark}

\begin{remark}
    \label{remark:parallel_nlp}
    Once the sampled trial points $x_L^i$ are obtained, the attempted computation of the corresponding labels via NLP solvers is naturally amenable to parallelization, since each optimization problem is independent. Accordingly, parallel computing provides a practical means of mitigating the computational burden associated with offline training-data generation. However, the number of sample points required to maintain a fixed resolution generally grows exponentially with the dimension of $x_L$, whereas parallelization yields at most approximately linear speed-up with the number of workers. We emphasize that this curse-of-dimensionality issue is generally encountered by any numerical approach that seeks to approximate the mapping $\mathcal{X}_A \to \mathcal{U}$.
\end{remark}

\subsection{Comparison with existing learning-based NCA methods}
Our proposed approach differs from existing learning-based approaches for solving general static nonlinear control allocation problem. In particular, we focus on applying learning to address the feedback linearization errors described in \cref{sec:indi-issue-nonaffine}, which arise when applying standard INDI to input nonaffine systems. Previous work \cite{khan2024_nonlin_CA_learning_based_approach, huan2018_ECC_NCA_using_deep_AE_NN} does not generate labels using NLP. Instead, the primary training objective is to minimize the average allocation error over a collection of sampled points in the set $\mathcal{X}_g \times \mathcal{M}$, i.e., to minimize the average normed residual of \cref{eq:nca_nonaffine} where $u$ is replaced by the network output that depends on $x_0$ and $\mu_{des}$. In fact, in \cite{khan2024_nonlin_CA_learning_based_approach}, only the allocation error is minimized while in \cite{huan2018_ECC_NCA_using_deep_AE_NN} a joint loss function is proposed that allows the network to seek simultaneous minimization of the average normed allocation error and a secondary static control objective that depends on the output $\hat{u}^*$ of the network. We believe that our proposed method makes more efficient use of the training data than existing approaches. This is because our method explicitly separates the problem of approximating the inverse mapping at inter-sample points from the problem of enforcing allocation accuracy and optimality at the sampled training points. The latter is handled through direct solution of the corresponding NLPs, for which established solvers are used to generate high-quality labels in a computationally efficient manner with point-wise accuracy guarantees at the sampled points. The learning model is tasked solely with interpolation of the inverse map over the input space.

\section{Simulation results}\label{sec:example_and_results}

\begin{figure}[H]
    \centering
    \includegraphics[width=1\columnwidth]{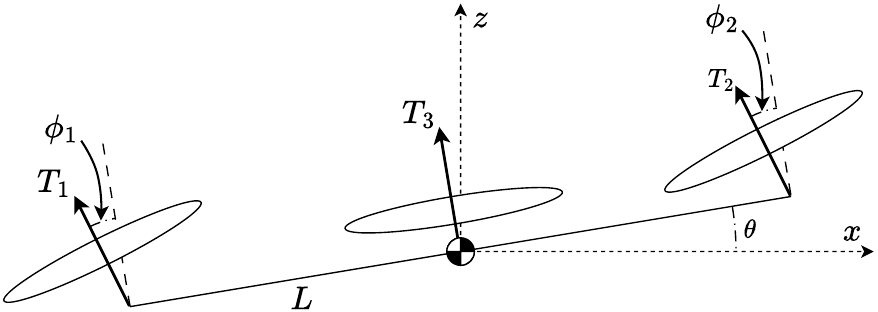} 
    \caption{Diagram of the conceptual planar bi-tilt tricopter}
    \label{fig:system_diagram}
\end{figure}

We consider a simple example system to: (1) demonstrate the issues described in \cref{sec:standard_indi_problems} associated with applying standard INDI with linearized CA to input nonaffine systems, and (2) to show the effectiveness of our proposed learning-based approach from \cref{sec:proposed_learning_approach} in solving the NCA problem that results in \cref{sec:indi_inspired_nonaffine} after explicitly avoiding the issues with the standard formulation. The example system is a conceptual planar bi-tilt tricopter. A diagram of the system is shown in \cref{fig:system_diagram}. Rotors 1 and 2 at the left and right ends, respectively, provide thrust inputs $T_1$ and $T_2$ and are capable of tilting about axes normal to the $x$-$z$ plane. The tilt angles $\phi_1$ and $\phi_2$ are measured as shown in \cref{fig:system_diagram}, where a tilt to the left is taken as positive by convention. Rotor 3, located at the vehicle center, provides thrust input $T_3$ and has a fixed orientation in the vehicle frame, i.e., rotor 3 cannot be tilted. The state vector $x = [v_x, v_z, \dot{\theta}, \theta ]^T \in \mathbb{R}^4$ consists of the linear velocities $v_x$ and $v_z$ of the vehicle in the $x$ and $z$ directions, along with the angular velocity of the vehicle, $\dot{\theta}$, and the attitude angle, $\theta$. The input vector is given by $u = [T_1, T_2, T_3, \phi_1, \phi_2]^T \in \mathbb{R}^5$.
The simplified system dynamics obey \cref{eq:nonaffine_dyn}, with the right-hand side, $f(x)+g(x,u)$, given by 
\vspace{-1mm}
{\small  
\begin{equation}  \label{eq:bi_tilt_tricopt_dyn}
 \hspace{-1mm}   
    \begin{bmatrix}
        0 \\ -g \\ 0 \\ x_3
    \end{bmatrix} \!+\!
  \begin{bmatrix}
        - \frac{1}{m} (u_3 \sin (x_4) \!+\! u_1 \sin (x_4\!+\!u_4) \!+\! u_2 \sin(x_4 \!+\! u_5)) \\ \frac{1}{m}(u_3 \cos(x_4) \!+\! u_1 \cos(x_4 \!+\! u_4) \!+\! u_2 \cos(x_4 \!+\! u_5)) \\ \frac{L}{I_y}(-u_1 \cos (u_4) \!+\! u_2 \cos (u_5)) \\ 0 
    \end{bmatrix},\!
\end{equation}}where $m$ is the vehicle mass, $I_y$ is the moment of inertia, $L$ is the arm length, and $g$ is the acceleration due to gravity. We consider box constraints on the inputs of the form $\underline{u} \leq u \leq \overline{u}$ with $0 \leq T_i \leq mg$ for each of the three rotors and $-60^{\degree} \leq \phi_i \leq 60^\degree$ for both tilt angles.

\begin{figure}[h]
    \centering
    \includegraphics[width=1\columnwidth]{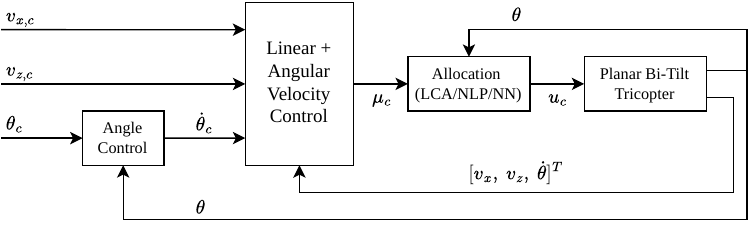} 
    \caption{Control architecture for the planar bi-tilt tricopter}
    \label{fig:controller_architecture}
\end{figure}

We design a controller to track linear velocity and attitude reference commands. Attitude control is achieved through a cascaded structure. The overall control architecture is shown in \cref{fig:controller_architecture}.
The angle controller is designed using proportional control: $\dot{\theta}_c = -K_{\theta}(\theta - \theta_c)$, with $K_{\theta} = \frac{1}{0.3}$ to achieve a target time constant of $0.3$s for the attitude control loop. The inner linear + angular velocity control loop is designed using feedback linearization. Let $x_r$ denote the reduced state consisting of $[v_x, v_z, \dot{\theta}]^T$. The virtual control $\nu(x_r)$, which represents the desired dynamics $\dot{x}_{r, des}$, is designed using proportional control: $\nu(x_r) = -K(x_r - x_{r,c})$ where $x_{r,c}$ is the commanded signal for $x_r$ and $K = \text{diag}(K_{v_x}, K_{v_z}, K_{\dot{\theta}} )$. We select $K_{v_x} = \frac{1}{0.3}$, $K_{v_z} = \frac{1}{0.4}$, and $K_{\dot{\theta}} = \frac{1}{0.03}$. The gain for angular rate control is selected such that the time constant for desired angular rate response is separated from that of the desired angle response by a factor of 10 to ensure adequate bandwidth separation of the angle control cascade. The gain for $v_z$ control is selected to be similar to what is commonly achievable in small multirotor drones. Typically, the time constant for the lateral velocity ($v_x$, in our case) response of multirotor drones is slower than that of the vertical velocity response due to the fact that non tilt-rotor drones must first change their attitude in order to generate a lateral force. We target a more aggressive lateral velocity response because of the ability to directly generate lateral forces that is provided by the rotor tilt capability.

For the sake of discussing the linear + angular velocity control via feedback linearization in the context of our developments in previous sections, we can consider the first three rows of \cref{eq:bi_tilt_tricopt_dyn} as the dynamics \cref{eq:nonaffine_dyn}, i.e., we can consider the feedback linearization of the reduced state $x_r$ where the dynamics are given by $\dot{x}_r = f_r(x_r) + g_r(\theta, u)$ with $\theta$ representing $x_g$ from \cref{sec:proposed_learning_approach}. Note that the drift term $f_r(x)$ is constant, which helps isolate the effect of the approximation of the nonlinear actuator model $g_r(\theta, u)$.
We compare three design methods for attempting to achieve feedback linearization of the inner loop:
\begin{enumerate}
    \item Standard INDI with LCA via local linearization of $g_r(\theta,u)$
    \item Nonaffine-INDI (\cref{sec:indi_inspired_nonaffine}) with NCA solved using online NLP.
    \item Nonaffine-INDI (\cref{sec:indi_inspired_nonaffine}) with NCA solved using the proposed learning-based approach from \cref{sec:proposed_learning_approach}.
\end{enumerate}

For brevity, we will refer to the methods 1), 2), and 3) above as LCA, NLP, and NN, respectively. For the LCA method, we use a QP approach similar to \cref{eq:QP_alloc_general_form} with inequality constraints (actuator limits) added to solve the linear allocation problem \cref{eq:linear_incremental_CA} with $B_0$ computed via analytical Jacobian. However, we minimize a quadratic form of the actual input $u$ rather than of $\Delta u$: $(u_0 + \Delta u)^T Q (u_0 + \Delta u) = \Delta u ^T Q \Delta u + 2u_0^T Q \Delta u$ where the equality holds for symmetric $Q$. For the NLP method, we minimize the same secondary objective $J(u) = u^T Qu$ subject to the nonlinear condition \cref{eq:nca_nonaffine}. We use MATLAB \texttt{quadprog} with the `interior-point-convex' algorithm to solve the QP problem and MATLAB \texttt{fmincon} with the `interior-point' algorithm to solve the NLP problem. For online solving of the NLP, we use the solution from the previous time step as an initial guess to warm-start the NLP solver. For the NN method, we obtained the data $X$ using the heuristic method proposed in \cref{sec:proposed_learning_approach} and solved the NLP for each sample point with the same method as was used for online evaluation. After rejecting infeasible sample points, we retained $N = 118,936$ samples for training the network. The network consisted of three hidden layers with 128 neurons per layer and $\tanh$ activations. Our network architecture was chosen arbitrarily. Proper ablation studies could be used to find an appropriate balance between network complexity, inference time, and prediction accuracy for a particular problem. Our aim was simply to demonstrate the proposed method.

To handle infeasible allocation problems in either the LCA or NLP methods, we utilized a prioritized relaxation scheme, as described in \cref{remark:ca_relaxation}, that was triggered in the event of unsuccessful exit flags of either \texttt{quadprog} or \texttt{fmincon}.

We simulated the system in response to two reference trajectories, with states and inputs initialized at the trim point $\tilde{x} = [0, 0, 0, 0]^T, \tilde{u} = [\frac{mg}{3}, \frac{mg}{3}, \frac{mg}{3}, 0, 0]^T$. The first reference trajectory, denoted by \textit{SingleStep}, consists of a step command of $+3$ m/s in $v_x$ at $t = 1$s with the $v_z$ and $\theta$ commands held at 0 over the $15$s trajectory duration. The second reference trajectory, denoted by \textit{TripleDoublet}, consists of three sequential doublets, one for each controlled variable, in the sequence $v_x$, $v_z$, $\theta$ with the trajectory duration also being $15$s.

\begin{figure}[h]
\begin{subfigure}{1\columnwidth}
    \centering
    \includegraphics[width=1\columnwidth]{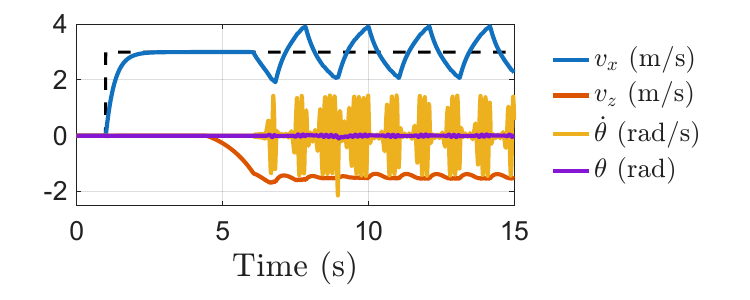}
    \vspace{-15mm}
\end{subfigure}
\begin{subfigure}{1\columnwidth}
    \centering
    \includegraphics[width=1\columnwidth]{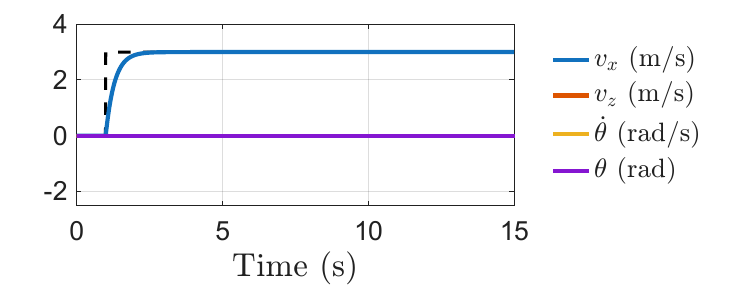}
     \vspace{-15mm}
\end{subfigure}
\begin{subfigure}{1\columnwidth}
    \centering
    \includegraphics[width=1\columnwidth]{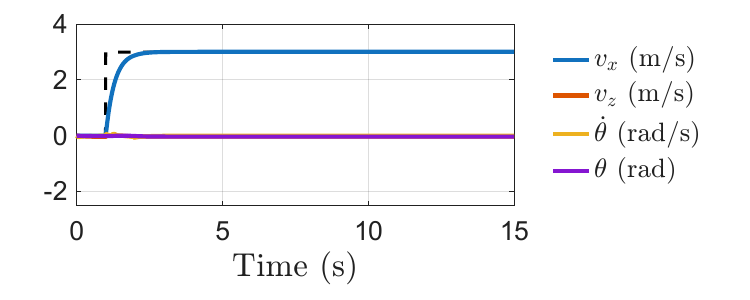}
\end{subfigure}
\caption{State trajectories for the \textit{SingleStep} reference trajectory under the LCA (top), NLP (middle), and NN (bottom) methods.}
\label{fig:SingleStep_plots}
\end{figure}

\begin{figure}[h]
\centering
    \includegraphics[width=1\columnwidth]{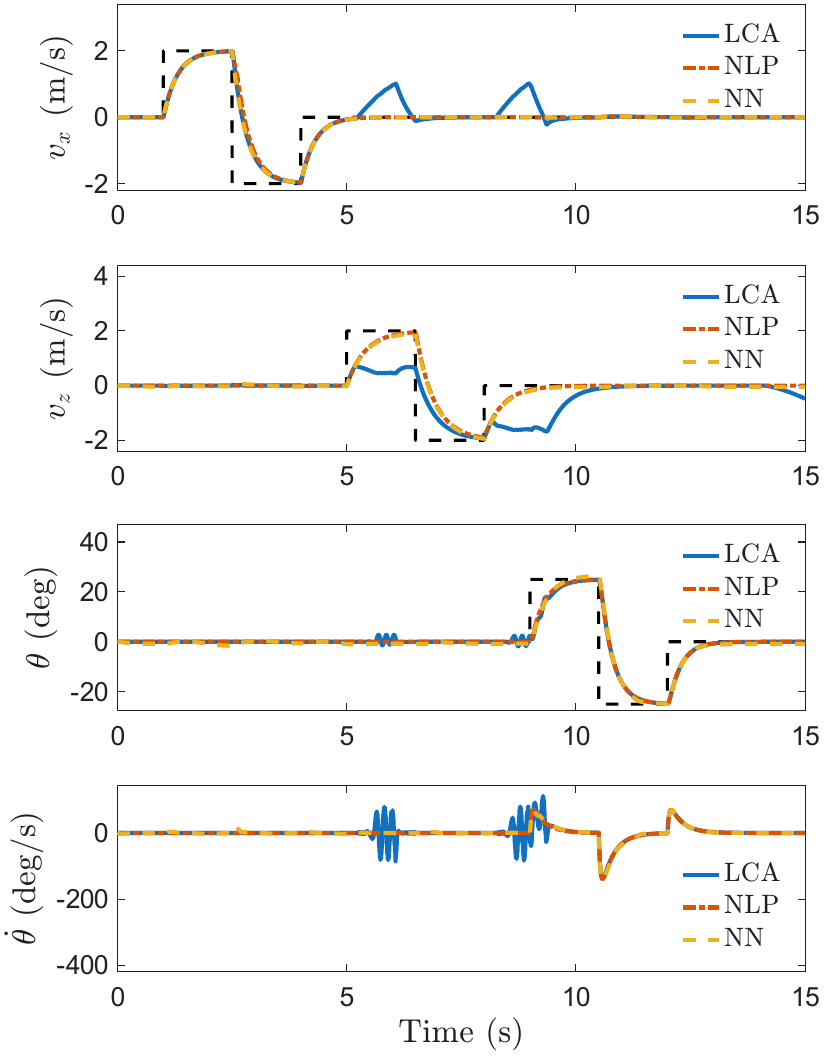}
    \vspace{-4mm}
\caption{State trajectories for the \textit{TripleDoublet} reference trajectory under the LCA, NLP, and NN methods.}
\label{fig:TripleDoublet_plots}
\end{figure}

The responses to the \textit{SingleStep} reference trajectory are shown in \cref{fig:SingleStep_plots}. The LCA method shows good tracking initially, but encounters a severe performance degradation starting from approximately $t=5$s characterized by high-frequency oscillations. The NLP method achieves the ideal response, while the NN method yields an almost identical response.
The response to the \textit{TripleDoublet} reference trajectory is shown in \cref{fig:TripleDoublet_plots}, with individual state components shown in each subplot. The responses to the first and third doublets are similar for all three methods, but the LCA method results in a poor response to the second doublet corresponding to the vertical velocity. The NN method achieves similar performance to the NLP method, as was the case for the \textit{SingleStep} reference trajectory.

\begin{figure}[h]
\centering
    \includegraphics[width=1\columnwidth]{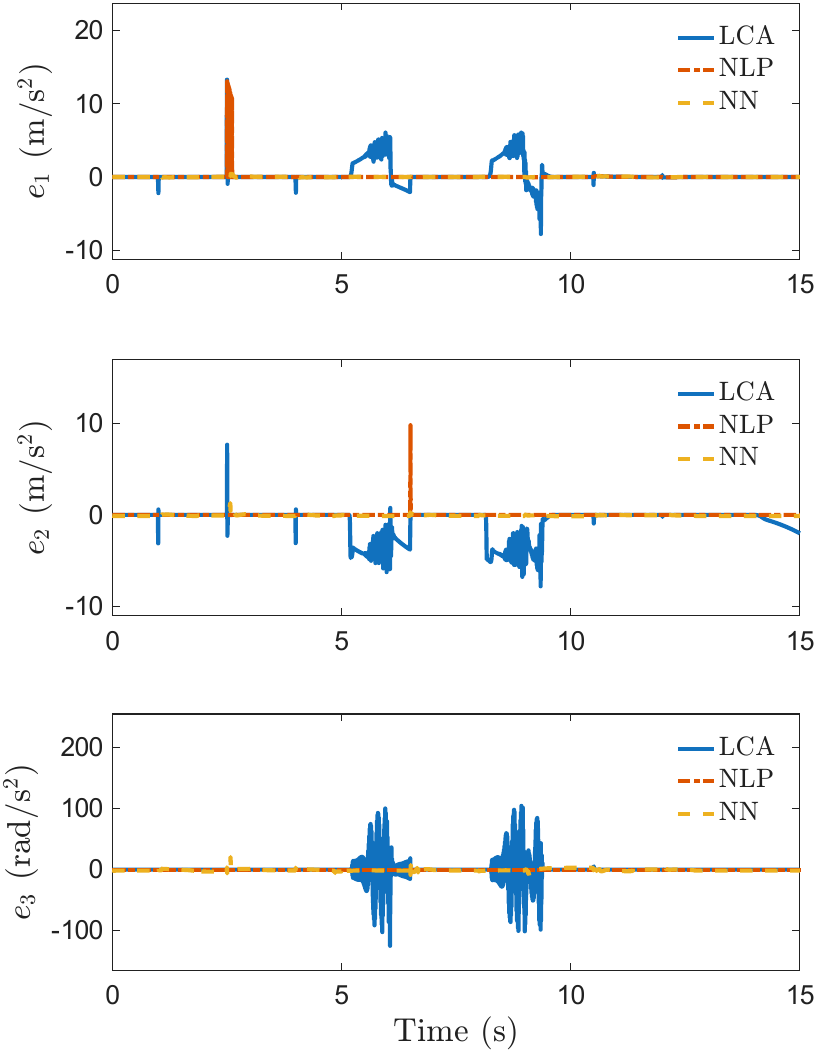} 
\caption{Feedback linearization errors, defined as $e = \dot{x} - \nu(x)$, for the \textit{TripleDoublet} reference trajectory under the LCA, NLP, and NN methods.}
\label{fig:TripleDoublet_fbl_errs}
\vspace{-4mm}
\end{figure}

The feedback linearization error for the \textit{TripleDoublet} reference trajectory is shown in \cref{fig:TripleDoublet_fbl_errs}. Note that the spikes in the error under the NLP and NN methods at the halfway points of the first two doublets are due to temporarily infeasible allocation problems under the input constraints. However, the LCA method results in persistently large feedback linearization errors during the second doublet, even when the QP remains feasible under the linearized equality constraint \cref{eq:linear_incremental_CA}. The LCA method also experiences spikes in allocation error during the step changes within the first doublet. This provides direct evidence of the problem with applying standard INDI + linear CA to nonaffine control systems that was discussed in \cref{sec:standard_indi_problems}. 

A comparison of allocator computational times is provided in \cref{tab:comp_time}. The times listed for the LCA method corresponds solely to the solution time of the QP problem. The local effectiveness matrix $B_0$ was computed analytically in our case. In practice, one may need to compute $B_0$ using online numerical differentiation or obtain $B_0$ from look-up tables, etc. The times listed for the NLP method correspond solely to the solution time of the NLP problem. The times listed for the NN method include only the network inference time. This does not account for pre- and post-processing (normalization) time. However, these operations generally involve only simple arithmetic operation and thus would likely not contribute meaningfully to the overall inference time. The measurements in \cref{tab:comp_time} were obtained using the MATLAB \texttt{timeit} function, which is more robust than the \texttt{tic-toc} timing method. The measurements were conducted on a Windows machine with an Intel Core i7-14700 processor, which is not representative of the typical computing unit on common UAVs, for instance. However, we expect that the relative differences would approximately hold on different hardware.
\begin{table}[h]
\centering
\caption{Allocator Computational Times}
\label{tab:comp_time}
\resizebox{\columnwidth}{!}{%
\begin{tabular}{c|cc|}
\cline{2-3}
\multicolumn{1}{l|}{} & \multicolumn{1}{c|}{\textit{SingleStep}} & 
\textit{TripleDoublet} \\
\cline{2-3}
\multicolumn{1}{l|}{} & 
\multicolumn{2}{c|}{Mean $\pm$ SD [Max] (ms)} \\
\hline
\multicolumn{1}{|c|}{LCA} & \multicolumn{1}{c|}
{$0.84 \pm 0.03 \ [1.29]$} &
$0.85 \pm 0.06 \ [2.50]$ \\
\hline
\multicolumn{1}{|c|}{NLP} & \multicolumn{1}{c|}
{$3.22 \pm 3.42 \ [81.7]$} &
$5.78 \pm 18.37 \ [744]$ \\
\hline
\multicolumn{1}{|c|}{NN}  & \multicolumn{1}{l|}
{$0.69 \pm 0.04 \ [1.02]$} & \multicolumn{1}{l|}
{$0.71 \pm 0.08 \ [2.59]$} \\ \hline
\end{tabular}%
}
\end{table}

The results in \cref{tab:comp_time} indicate that both LCA and NN enjoy small average runtime with a stable distribution, with the NN method performing slightly faster in general than the LCA method. The average runtime of the NLP allocator is substantially higher than that of the LCA and NN methods, though the average is on roughly the same order of magnitude. Importantly, the known flaws of the NLP method are highlighted by the standard deviation (SD) and worst-case runtimes, which indicate that NLP with no modification is not suitable for real-time implementation. For this particular example, the NN method achieved nearly identical performance to the NLP method for the reference trajectories that we tested. The NN method achieved this near-ideal response and outperformed LCA while actually being more computationally efficient than LCA in terms of online evaluation time. 
We do not claim that this will be uniformly true for all possible reference trajectories.

\section{Conclusion}\label{sec:conclusion}

In this paper, we identified and described a fundamental problem concerning the application of standard INDI + linearized CA to input nonaffine systems. We showed that to avoid this problem while still using static CA, the solution of an NCA problem is required. We proposed a novel learning-based approach to provide approximate online solutions to the NCA problem with efficient evaluation. Numerical simulations validated the issues with linearized CA for input nonaffine systems and demonstrated the effectiveness of the proposed learning-based approach. Future work will aim to develop an effective hybrid NLP + learning-based approach and apply these techniques to more complex systems. Additionally, we plan to investigate the use of similar methods for fault-tolerant control of hybrid VTOL aircraft. 

\section*{Acknowledgment}
Generative AI tools were used to help polish the language in certain sections of the manuscript.

\bibliographystyle{ieeetr}
\bibliography{bib/refs-adam}

\end{document}